\documentstyle[11pt,COIMBRA_pasp,epsf]{article}      

\begin{document}

\title{Significant discrepancies between stellar evolution models and
solar-type eclipsing and visual binaries ? }

\author{E. Lastennet}
\affil{Depto. de Astronomia, UFRJ, Rio de Janeiro, Brazil}

\author{J. Fernandes}
\affil{Observat\'orio Astron\'omico da Univ. de Coimbra, Portugal}

\author{E. Oblak}
\affil{Observatoire de Besan\c{c}on, UMR CNRS 6091, France}

\author{D. Valls-Gabaud}
\affil{UMR CNRS 5572, Observatoire Midi-Pyren\'ees, Toulouse, France}

\begin{abstract}
Individual components of well-detached binary systems are assumed
to be two single-like stars with a common origin, i.e. they share the
same chemical composition and same age.
Therefore, one expects to fit the observed parameters of both components
with a single isochrone at the same metallicity.
We show that serious problems appear for systems with accurate
fundamental data (eclipsing binaries in the field and in the Hyades,
and a visual binary) in the 0.7-1.1 M$_{\odot}$ mass range.
We discuss and briefly review the results obtained so far on these objects.
Finally, in an attempt to solve this problem, we present new projects, both
on the theoretical and the observational sides.

\end{abstract}



\section{Introduction to the problem}

The study of stars with masses larger than $\sim$0.6 M$_{\odot}$
bypasses difficulties in the treatment of the equation of state and
the atmosphere.
On the other hand,
stars with masses greater than about 1.1 M$_{\odot}$ have a permanent convective
core, introducing an additional parameter, the amount of overshooting.
Therefore, we expect current stellar evolution models to be able to match the
basic properties of stars in the 0.7-1.1 M$_{\odot}$ mass range.
In particular, some detached binaries provide very accurate stellar data so they are
ideal candidates to critically test sets of theoretical models.
Nevertheless, as pointed out by Popper (1997),
 {\it a serious dilemma appears
 to be present in the comparison of fundamental stellar properties derived
 from observations and the predictions of stellar models.}
In the next sections we review stellar objects
(EBs in the field and in the Hyades and one visual binary) suggesting
that current stellar models present some problems around the mass of the Sun.

\section{Eclipsing binaries in the field}

We consider the following detached eclipsing binaries (DEBs):
RT And, HS Aur, CG Cyg and FL Lyr.
We obtain bad fits
by fitting both components
simultaneously in the HR diagram with Padova group (Fagotto et al. 1994)
or Geneva group (Mowlavi et al. 1998) models.
Either these stellar models are unreliable
in this part of the HRD, or the T$_{\rm eff}$s of (at least) the
secondaries need revision.
As shown by Pols et al. (1997),
the same difficulty appears with the Cambridge group models
by fitting simultaneously the effective temperature (T$_{\rm eff}$),
mass ($M$) and radius ($R$).
Clausen et al. (1999) recently reviewed this dilemma with other sets of
theoretical models, getting bad fits as well. \\
Could the measurements be responsible of this problem ?
The relative error on the individual T$_{\rm eff}$s of the sample
is always small ($<$3.5\%). However, while the mass and radius of each
component of DEBs can be accurately measured (1-2\%),
the use of T$_{\rm eff}$  (and so luminosity) is not as reliable because
derived indirectly (from various photometric or spectroscopic indicators) and
thus may explain  part of the problem.
Lastennet et al. (1999a), Ribas et al. (2000), and Lastennet, Cuisinier \&
Lejeune (these proceedings) attempted to carefully re-derive reliable
EBs T$_{\rm eff}$s,
but unfortunately none of these works study
the stars in question
because individual $uvby$ photometry would be necessary.
Mass transfer should not be an explanation because
none of these stars overflows its Roche lobe (Lastennet 1998):
HS Aur A: $\sim$11\%, B: $\sim$10\%;
FL Lyr A: $\sim$35\%, B: $\sim$29\%;
CG Cyg A: $\sim$61\%, B: $\sim$60\%;
RT And A: $\sim$77\%, B: $\sim$64\%.
Nevertheless, for the system RT And,
the face-to-face position of the spots on the surface of both
components
may indicate the possibility of a mass transfer from the
primary to the secondary component through a magnetic bridge connecting both
active regions (Pribulla et al. 2000).

\section{Eclipsing binaries in open clusters}

As discussed for instance by Lastennet et al. (2000),
DEBs
members of a star cluster provide stringent constraints
on stellar evolution models when the metallicity of the cluster is
known.
The eclipsing binary V818 Tauri (a member of the Hyades) contains 2
stars in the 0.7-1.1 M$_{\odot}$ mass range (1.072$+$0.769 M$_{\odot}$,
Peterson \& Solensky 1988).
Once again, a similar analysis for this binary (Lastennet et al.
1999b) found no Padova models fitting simultaneously $M$ and $R$
(thus, without
using any information on T$_{\rm eff}$) for any age or metallicity.

\section{Visual binaries}

Such a problem was also detected in this mass range by Fernandes et al. (1998)
for 85 Peg (0.91$\pm$0.11 M$_{\odot}$$+$0.73$\pm$0.13 M$_{\odot}$),
a nearby visual binary star ($\sim$12.4 pc).
They found no solution satisfying the constraint of the
luminosity ($L$) and T$_{\rm eff}$ for both stars and corresponding to
the observed metallicity and sum of the masses.
85 Peg  A and B appeared to be too cold and/or over-luminous with
respect to the ZAMS.
Only models with extremely low helium ($Y$$<$0.20) and high age (20
Gyr) could fit the HR diagram position of 85 Peg A, which seems definitively
unrealistic since the primordial helium is estimated to be
$Y$$=$0.232$\pm$0.003 (Olive \& Steigman 1995)
and the age of the Universe to be
between 10-20 Gyr.
They succeeded to fit the 85 Peg A position by decreasing
$\alpha$$_{MLT}$  to about 1.0
but a similar change of $\alpha$$_{MLT}$ didn't fit
the 85 Peg B position.
A more recent study (Lebreton et al. 1999) shows that including the
diffusion, the contribution of the $\alpha$-elements, and non LTE-effects
in low metallicity models help solving the above discrepancy, but
there is still a problem to match the mass of the secondary.
A new study of 85 Peg is underway (Fernandes et al., in prep.)
but the influence of rotation can already be excluded because vsin$i$ is very low
(vsin$i$$<$5 km.s$^{-1}$, according to the catalog of Glebocki \&
Stawikowski, 2000).

\section{Future theoretical projects with CESAM}

The systematic discrepancy presented previously is not observed in CD Tau C,
a solar mass companion of the triple system CD Tau: Ribas et al. (1999)
obtained a perfect fit of the 3 components with a single isochrone.
Therefore, except CD Tau C (whose mass still needs an accurate measure),
the other examples show that there is still great problems for the
stellar evolution theory to predict the properties of solar-type stars.
However, this is of first importance for stellar theory to match at least
the best known objects (the Sun and non-interacting binary systems)
before any attempt to derive information for star clusters or
stellar populations in galaxies.
To tackle this problem, the CESAM code (Morel 1997)
should be very useful
to explore the influence of each physical parameter.
CESAM
performs calculations of
1D quasi-static stellar evolution  including diffusion and mass-loss
and computes the evolution of stars from the pre-main
sequence (PMS) to the beginning of the red giant branch.

As suggested by Clausen et al. (1999), the problem briefly presented
in the previous sections may be removed (or
at least diminished) if a significantly lower $\alpha_{MLT}$ is adopted
for the less massive secondary components.
A recent study on V818 Tau (Lebreton et al. 2001)
seem to support this idea and similar
tests on the EBs of \S2 are in progress (Lastennet \& Fernandes).
We suggest
a detailed study of each system to check if there is a solution
for a set of ($\alpha_{MLT}$, $Y$, $Z$, age).
Even if the effect of stellar rotation is usually negligible for stars
less massive than $\sim$1.4 M$_{\odot}$, an even slightly effect on
T$_{\rm eff}$ and $L$
may be relevant for some of these stars
(e.g. vsin$i$ $\sim$70 km.s$^{-1}$ for both components of CG Cyg).
Another explanation may be that some of these stars are still in the
PMS phase, and this is another advantage to use the CESAM code which
includes this phase.

\section{Future constraints from new observational campaigns and GAIA}

Of course, the number of relevant systems is still very small to give a
definitive conclusion.
A photometric and spectroscopic observational campaign of about 50
newly discovered Hipparcos EBs at OHP (France), Kryonerion
(Greece)  and Cracow (Poland) would increase the number of objects
(Kurpinska-Winiarska \& Oblak 2000), as well as the dedicated observational
program of late F, G and K type stars  at the Danish 50 cm SAT (La Silla Obs.,
Chile) which already provided  new candidates (Clausen, Helt \& Olsen 1999).
The GAIA satellite would also increase the number of stars with individual
masses.
Halbwachs \& Arenou (1999) estimated that the individual masses of 79
double-lined spectroscopic binaries (SB2s) - which are not eclipsing binaries
- in the {8$^{th}$ catalogue of orbital elements of SB systems} (Batten
et al. 1989) would be derived from the GAIA astrometric observations.
We expect some of them ($\sim$ 10 according to Tab. 1 of Halbwachs \& Arenou)
to belong to the mass range discussed in the present work.

\section{Conclusion}

We review solar-type stars showing a clear discrepancy between
their
fundamental properties and stellar model predictions.
This indicates problems in current stellar evolutionary models
because a revision of their T$_{\rm eff}$ would not solve the
disagreement regarding the masses and radii (see Tab. 7 of Popper
1997, or Fig. 4 in Lastennet et al. 1999b).
These stars belong to binary systems without interaction so representative
of single stars, and it is of first importance for stellar theory to match
at least the best known objects (the Sun and non-interacting binary systems).
In an attempt to solve this problem, we suggest some theoretical
explorations, in particular a detailed study of stellar rotation and PMS
phases.
We also present some projects which should help to increase the limited
number of accurate fundamental parameters of detached EBs.

\acknowledgments
E.L. thanks J.-L. Halbwachs for useful
comments about SB2s and GAIA.


\small
\begin{references}

\reference Clausen J.V., et al., 1999, ASP Conf. Series 173, 265
\reference Clausen J.V., Helt B.E., Olsen E.H., 1999, ASP Conf. Series 173, 321
\reference Fagotto F., Bressan A., Bertelli G., Chiosi C., 1994c, \aaps, 105,
39
\reference Fernandes J., Lebreton Y., Baglin A., Morel P., 1998,
\aap, 338, 455
\reference Glebocki R., Stawikowski A., 2000, Acta Astronomica, 50, 509.
\reference Halbwachs J.L., Arenou F., 1999, Baltic Astronomy 8, 301
\reference Kurpinska
M., Oblak E., 2000, IAU Symp. 200,
eds. B. Reipurth \& H. Zinnecker, p.141
\reference Lastennet E., 1998, Ph-D  thesis, Univ. L. Pasteur, Obs. de Strasbourg, France
\reference Lastennet E., Lejeune Th., Westera P., Buser R., 1999a, \aap, 341,
857
\reference Lastennet E., Valls-Gabaud D., Oblak E., 2000,
IAU Symp. 200,
eds. B. Reipurth \& H. Zinnecker, p.164
\reference Lastennet E., Valls-Gabaud D., Lejeune Th., Oblak E., 1999b, 
\aap, 349, 485
\reference Lebreton Y., et al., 1999, \aap, 350, 587
\reference Lebreton Y., Fernandes J., Lejeune Th., 2001, A\&A accepted
(astro-ph/0105497)
\reference Morel P., 1997, \aaps, 124, 597
\reference Mowlavi  N. et al.,
1998, \aaps, 128, 471
\reference Olive K.A., Steigman G., 1995, \apjs, 97, 49
\reference Peterson D.M.,  Solensky R., 1988, \apj, 333, 256
\reference Pols O.R. et al.,
1997, \mnras, 289, 869
\reference Popper D.M., 1997, \aj, 114, 1195
\reference Pribulla T. et al.,
2000, \aap, 362, 169
\reference Ribas I. et al., 1999, \mnras, 309, 199
\reference Ribas I., Jordi C., Torra J., Gim\'enez A., 2000, \mnras, 313, 99

\end{references}
\end{document}